\documentstyle[prb,aps,psfig]{revtex}

\begin{document}

\draft

\title{Electron transport through a circular constriction} 
\author{Branislav Nikoli\' c and Philip B. Allen} 
\address{Department of Physics and Astronomy,
SUNY at Stony Brook, Stony Brook, New York 11794-3800} 
\maketitle

\begin{abstract} 
We calculate the conductance of a circular  constriction of radius 
$a$ in an insulating diaphragm which separates two conducting 
half-spaces characterized by the mean free path $\ell$. Using the 
Boltzmann equation we obtain an answer for all values of the ratio 
$\ell/a$. Our exact result interpolates between the Maxwell 
conductance in diffusive ($\ell \ll a$) and the Sharvin conductance in 
ballistic ($\ell \gg a$) transport regime. Following the
earlier approach of Wexler we find the explicit form of the Green's 
function for the linearized Boltzmann operator. 
The formula for the conductance deviates by less than $11\%$ from the 
naive interpolation formula obtained by adding resistances in the 
diffusive and the ballistic regime.
\end{abstract}


\section{Introduction}
\begin{figure}
\centerline{
\psfig{file=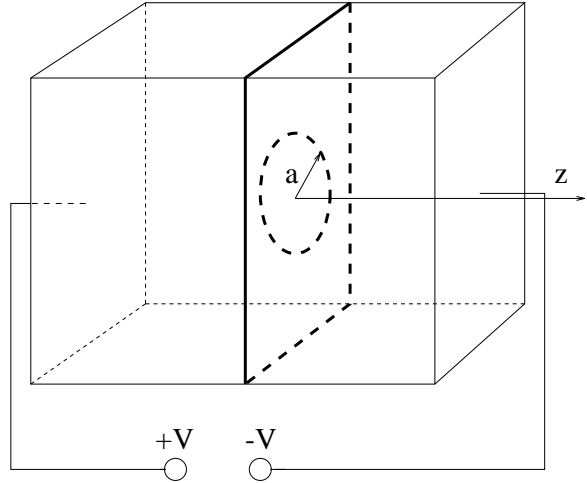,height=2.5in,angle=-90}}
\vspace{0.2in}
\caption{Electron transport through the circular constriction in an 
insulating diaphragm separating two conducting half-spaces (each with 
a mean free path $\ell$).}
\label{fig:fig1}
\end{figure}

The problem of electron transport through an orifice (also known as a point
contact) in an insulating diaphragm separating two large conductors
(Fig.~\ref{fig:fig1}) has been studied for more than a century.
Maxwell~\cite{maxwell} found the resistance in the diffusive regime when the
characteristic dimension $a$ (radius of the orifice) is much larger than the
mean free path $\ell$. Maxwell's answer, obtained from the solution of Poisson
equation and Ohm's law, is
\begin{equation}
\label{eq:max}
R_M=\frac{\rho}{2a},
\end{equation}
where $\rho$ is resistivity of the conductor on each side of the diaphragm.
Later on, Sharvin~\cite{shar} calculated the resistance in the ballistic regime
($\ell \gg a$)
\begin{equation}
\label{eq:shar}
R_S=\frac{4 \rho \ell}{3 A}=\left(\frac{2e^2}{h}\frac{k_F^2 A}{4 \pi}
\right)^{-1},
\end{equation}
where A is the area of the orifice.
This ``contact resistance'' persists even for the ideal conductors (no scattering) 
and has a purely geometrical origin, because only a finite current can flow through 
a finite size orifice for a given voltage. In the Landauer-B\" uttiker transmission 
formalism~\cite{lb}, we can think of a reflection when a large number of transverse 
propagating modes in the reservoirs matches a small number of propagating modes in 
the orifice. In the intermediate regime, when $a \simeq \ell$, the crossover 
from $R_M$ to $R_S$ was studied by Wexler~\cite{wex} using the Boltzmann equation 
in a relaxation time approximation. The influence of inelastic collisions on the orifice
current-voltage characteristics was studied using classical kinetic equations in
Ref.~\onlinecite{rus} and quantum kinetic equations (Keldysh formalism) in
Ref.~\onlinecite{rus1}. This effect underlies an experimental technique for the
extraction of the phonon density of states from the nonlinear current-voltage
characteristics (point contact spectroscopy~\cite{jen}).  Recently, the size of
orifice has been shrunk to $a \simeq \lambda_F$ allowing the observation of
quantum-size effects on the conductance~\cite{wees,wharam}. In the case of a
tapered orifice on each side of a short constriction between reservoirs,
discrete transverse states (``quantum channels'') below the Fermi energy which
can propagate through the orifice give rise to a quantum version of
Eq.~(\ref{eq:shar}).  The quantum point contact conductance is equal to an integer
number of conductance quanta $2e^2/h$.

Here we report a semiclassical treatment using the Boltzmann equation. Bloch-wave
propagation and Fermi-Dirac statistics are included, but quantum interference 
effects are
neglected.  Electrons are scattered specularly and elastically at the diaphragm
separating the electrodes made of material with a spherical Fermi
surface. Collisions are taken into account through the mean free path $\ell$. A
peculiar feature is that the driving force can change rapidly on the length
scale of a mean free path around the orifice region. The local current density
depends on the driving force at all other points.
Our approach follows Wexler's~\cite{wex} study. We find an explicit form of the
Green's function for the integro-differential Boltzmann operator. The Green's
function becomes the kernel of an integral equation defined on the compact
domain of orifice. Solution of this integral equation gives the deviation from
the equilibrium distribution function on the orifice. Therefore, it defines the 
current through the orifice and its resistance.

\begin{figure}
\centerline{
\psfig{file=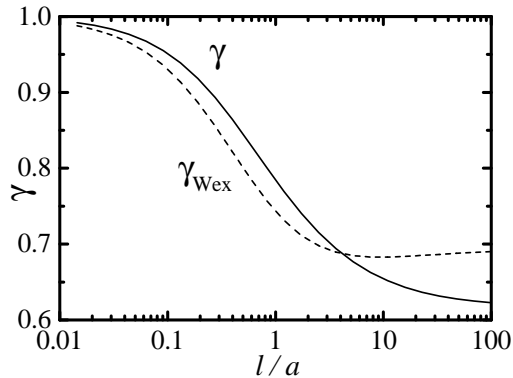,height=4.0in,angle=0}}
\vspace{-1.5in}
\caption{The dependence of factor $\gamma$ in Eq.~(\ref{eq:gamma}) on the 
ratio $\ell/a$. Also shown is the variational calculation of 
$\gamma_{\mathrm{Wex}}$ from Ref. 4.}
\label{fig:fig2}
\end{figure}

The exact answer can be written as
\begin{equation}
R(\ell/a) = R_S+\gamma(\ell/a)R_M,
\end{equation}
where $\gamma(\ell/a)$ has the limiting value $1$ as $\ell/a \rightarrow 0$ and
$R_S/R_M \rightarrow 0$. We are able to compute $\gamma(\ell/a)$ numerically to
an accuracy of better than $1\%$. Our calculation is shown on
Fig.~\ref{fig:fig2}. We also find the first order Pad\' e fit
\begin{equation}
\gamma_{\mathrm{fit}}(l/a)=\frac{1+0.83\, l/a}{1+1.33\, l/a},
\end{equation}
which is accurate to about $1\%$. Our answer for $\gamma$ differs little from
the approximate answer of Wexler~\cite{wex}, also shown on Fig.~\ref{fig:fig2}
as $\gamma_{\mathrm Wex}$.

Section~\ref{sec:semi} formulates the algebra and Sec.~\ref{sec:sol} explains
the solution.

\section{Semiclassical transport theory in the orifice geometry}
\label{sec:semi}
In order to find the current density ${\bf j}({\bf r})$ through orifice, in 
the semiclassical approach, we have to solve simultaneously the stationary 
Boltzmann equation in the presence of an electric field and the Poisson 
equation for the electric potential
\begin{eqnarray}
\dot{\bf r} \cdot \frac{\partial F({\bf k,r})}{\partial {\bf r}}-\frac{e\nabla
\Phi({\bf r})}{\hbar} \cdot \frac{\partial F({\bf k,r})}{\partial {\bf k}} & = &
- \frac{F({\bf k,r})-f_{\mathrm LE}({\bf k,r})}{\tau}, \label{eq:boltz}\\
\nabla^2 \Phi ({\bf r}) & = & - \frac{e\delta n({\bf r})}{\epsilon}, \\ \delta
n({\bf r}) & = &\frac{1}{\Omega} \sum_{\bf k}(F({\bf k,r})-f({\epsilon_k})), \\
0 & = & \frac{1}{\Omega} \sum_{\bf k} (F({\bf k,r})-f_{\mathrm LE}({\bf k,r})),
\\ {\bf j({\bf r})} & = & \frac{e}{\Omega} \sum_{\bf k}{\bf v_k}F({\bf k,r}).
\end{eqnarray}
Here $F({\bf k,r})$ is the distribution function, $f({\epsilon_k})$ is the
equilibrium Fermi-Dirac function, $\Phi({\bf r})$ is electric potential,
$\Omega$ is the volume of the sample and $f_{\mathrm LE}({\bf k,r})$ is a
Fermi-Dirac function with spatially varying chemical potential $\mu({\bf r})$
which has the same local charge density as $F({\bf k,r})$. In general, we have
to deal with the local deviation $\delta n({\bf r})$ of electron density from
its equilibrium value self-consistently. The collision integral is written in
the standard relaxation time approximation with scattering time
$\tau=l/v_F$. This system of equations should be supplemented with boundary
conditions on the left electrode (LE) at $z=-\infty$, right electrode (RE) at
$z=\infty$, and on the impermeable diaphragm $(D)$ at $z=0$:
\begin{mathletters}
\begin{eqnarray}
\label{boundary1}
\Phi({\bf r}_{\mathrm LE}) & = & V, \\
\label{boundary2}
\Phi({\bf r}_{\mathrm RE}) & = & -V, \\ j_z({\bf r}_D) & = & 0,
\end{eqnarray}
\end{mathletters}
where $z$-axis is taken to be perpendicular to the orifice.
In linear approximation we can express the distribution function $F({\bf k,r})$
and local equilibrium distribution function $f_{\mathrm LE}({\bf k,r})$ using
$\delta \mu ({\bf r})$ (local change of the chemical potential) and $\Psi ({\bf
k,r})$ (deviation function, i.e. energy shift of the altered distribution)
\begin{eqnarray} f_{\mathrm{LE}}({\bf k,r}) & = & f(\epsilon_k-\delta \mu ({\bf
r})) \approx f(\epsilon_k) -\frac{\partial f(\epsilon_k)}{\partial \epsilon_k}
\delta \mu ({\bf r}), \\ F({\bf k,r}) & = & f(\epsilon_k-\Psi ({\bf k,r}))
\approx f(\epsilon_k) -\frac{\partial f(\epsilon_k)}{\partial \epsilon_k} \Psi
({\bf k,r}).
\end{eqnarray}
These equations imply that $\delta \mu ({\bf r})$ is identical to the angular
average of $\Psi({\bf k,r})$
\begin{equation} 
\delta n({\bf r}) = \frac{1}{\Omega}\sum_{\bf k} -\frac{\partial
f(\epsilon_k)}{\partial \epsilon_k} \Psi({{\bf k,r}})=N(0)\langle\Psi({\bf
r})\rangle=N(0)\delta \mu({\bf r}),
\end{equation}
where $N(0)$ is the density of states at the Fermi energy $\epsilon_F$. In the
case of a spherical Fermi surface,
\begin{equation}
\langle \Psi({\bf r}) \rangle = \frac{1}{4\pi} \int d\Omega_k \Psi({\bf k,r}).
\end{equation}
Following Wexler~\cite{wex}, we introduce a function $u({\bf k,r})$ by writing
$\Psi({\bf k,r})$ as
\begin{equation}
\Psi({\bf k,r})=eVu({\bf k,r})-e\Phi({\bf r}).
\end{equation}
Thereby, the linearized Boltzmann equation~(\ref{eq:boltz}) becomes an
integro-differential equation for the function $u({\bf k,r})$
 
\begin{equation}
\label{eq:bu}
\tau {\bf v_k} \cdot \frac{\partial u({\bf k,r})}{\partial {\bf r}}=\langle
u({\bf r}) \rangle - u({\bf k,r}).
\end{equation}
To solve this equation we need to know only boundary conditions satisfied by
$u({\bf k,r})$ and then we can use this solution to find the potential
$\Phi({\bf r})$. Thus the calculation of the conductance from $u({\bf k,r})$ is
decoupled from the Poisson equation. This is an intrinsic property of linear
response theory~\cite{been}.  The boundary conditions for~(\ref{eq:bu}) are:
\begin{mathletters}
\begin{eqnarray}
\langle u({\bf r}_{\mathrm LE}) \rangle & = & 1, \\ \langle u({\bf r}_{\mathrm
RE}) \rangle & = & -1.
\end{eqnarray}
\end{mathletters}
They follow from the boundary conditions~(\ref{boundary1})-(\ref{boundary2}) for
the potential $\Phi({\bf r})$ and the fact that far away from the orifice we can
expect local charge neutrality entailing
\begin{equation}
\langle u({\bf r}) \rangle = \frac{\Phi({\bf r})}{V}.
\end{equation}
The driving force is explicitly absent from~(\ref{eq:bu}), but it enters the
problem through these boundary conditions. Since Eq.~(\ref{eq:bu}) is invariant
under the reflection in the plane of diaphragm
\begin{mathletters}
\begin{eqnarray}
({\bf k,r}) & \rightarrow & ({\bf k}^R,{\bf r}^R), \\ {\bf r}^R & = & (x,y,-z),
\\ {\bf k}^R & = & (k_x,k_y,-k_z),
\end{eqnarray}
\end{mathletters}
the boundary conditions imply that $u({\bf k,r})$ has reflection antisymmetry

\begin{equation}
u({\bf k,r})=-u({\bf k}^R,{\bf r}^R).
\end{equation}
Wexler's solution~\cite{wex} to the equation~(\ref{eq:bu}) relied on the
equivalence between the problem of orifice resistance and spreading resistance
of a disk electrode in the place of orifice. Technically this is achieved by
switching from the equation for function $u({\bf k,r})$ to the equation for
function
\begin{equation}
w({\bf k,r})=1+\mathrm{sgn}\,(z) u({\bf k,r}).
\end{equation}
The beauty of this transformation is that new function allows us to replace the
discontinuous behavior of $u({\bf k,r})$ on the diaphragm (which is the
mathematical formulation of specular scattering)
\begin{equation}
u({\bf k,r}_D-{\bf v_k}dt)=u({\bf k}^R,{\bf r}_D-{\bf v_k}dt)=-u({\bf k},
{\bf r}_D+{\bf v_k}dt)
\end{equation}
with continuous behavior of $w({\bf k,r})$ over the diaphragm, discontinuous
behavior over the orifice and simpler boundary conditions on the electrodes
\begin{equation}
\langle w({\bf r}_{\mathrm LE}) \rangle = \langle w({\bf r}_{\mathrm RE})
\rangle = 0.
\end{equation}
The Boltzmann equation~(\ref{eq:bu}) now becomes
\begin{equation}
\label{eq:bw}
\bbox{\ell}_{\bf k} \cdot \frac{\partial w({\bf k,r})}{\partial {\bf r}}+w({\bf
k,r})-\langle w({\bf r}) \rangle=s({\bf k,r})\delta(z)\theta(a-r),
\end{equation}
where we have introduced the function
\begin{equation}
s({\bf k,r})=2\ell_{\mathrm kz}u({\bf k,r})
\end{equation}
which is confined to the orifice region. It can be related to $w({\bf k,r})$ at
the orifice in the following way:
\begin{equation}
s({\bf k,r}_0)=2|\ell_{\mathrm kz}|(1-w({\bf k,r}_0-{\bf v_k}dt)).
\end{equation}
It plays the role of a ``source of particles'' in Eq.~(\ref{eq:bw}). The
notation ${\bf r}_0$ refers to a vector lying on the orifice, that is ${\bf
r}_0=(x,y,0)$ with $x^2+y^2 \leq a^2$. The discontinuity of $w({\bf k,r})$ on
the orifice is handled by replacing it by the disk electrode which spreads
particles into a scattering medium.

The Green's function for Eq.~(\ref{eq:bw}) is the inverse Boltzmann operator
(including boundary conditions)
\begin{equation}
\left (\bbox{\ell}_{\bf k} \cdot \frac{\partial}{\partial {\bf r}}+1-\hat{A} 
\right )
G_B({\bf k,r;k',r'})=\delta(\Omega_k-\Omega_{k'})\delta({\bf r-r'}),
\end{equation} 
and $\hat{A}$ is the angular average operator
\begin{equation}
\hat{A}f({\bf k})=\frac{1}{4\pi} \int d\Omega_k\, f({\bf k})=\langle f \rangle.
\end{equation}
The Green's function for the Boltzmann equation allows us to express $w({\bf
k,r}_0-{\bf v_k}dt)$ in the form of a four-dimensional integral equation over
the surface of the orifice
\begin{equation}
\label{eq:integralbw}
w({\bf k,r}_0-{\bf v_k}dt)=\int d\Omega_{k'}\,d{\bf r'}_0\,G_B({\bf k,r}_0-{\bf
v_k}dt;{\bf k',r'}_0+{\bf v_{k'}}dt)s({\bf k',r'}_0).
\end{equation}
The function $w({\bf k,r})$ is discontinuous over the orifice, so we formulate
the equation for this function at points infinitesimally close $(dt \rightarrow
+0)$ to the orifice.  We find the following explicit expression for the Green's 
function 
\begin{eqnarray}
G_B({\bf k,r;k',r'})=\frac{1}{\Omega}\sum_{\bf q} \frac{e^{i{\bf q \cdot
(r-r')}}}{1+i{\bf q \cdot} \bbox{\ell}_{\bf k}} \left(
\delta(\Omega_k-\Omega_{k'})+\frac{q\ell(q\ell-\arctan q\ell)^{-1}}{4\pi
(1+i{\bf q \cdot} \bbox{\ell}_{\bf k'})} \right ). \label{eq:gb}
\end{eqnarray}
Its form reflects the separable structure of Boltzmann operator, i.e. the sum 
of operators whose factors act in the space of functions of either $\bf r$ 
or $\bf k$. However it is nontrivial because the factors acting in 
$\bf k$-space do not commute and the Boltzmann operator is not normal---it 
does not have the complete set of eigenvectors and the standard procedure for 
constructing the Green's function from the projectors on these states fails. 
 The first term in~(\ref{eq:gb}) is singular and generates the
discontinuity of $w({\bf k,r})$ over the orifice.

\section{The Conductance of the orifice}
\label{sec:sol}
The conductance of the orifice is defined by
\begin{equation}
\label{cond}
G=\frac{1}{R}=\frac{I}{2V}=\frac{\int {d{\bf r}_0\, j_{z}({\bf r}_0)}}{2V},
\end{equation}
where the $z$-component of the current at the surface of the orifice is
\begin{equation}
j_{z}({\bf r}_0) = \frac{N(0)e^{2}V}{8 \pi \tau }\int{d{\Omega_{k}}\, s({\bf
k,r}_0)}.
\end{equation}
The Green's function result~(\ref{eq:gb}) allows us to rewrite
Eq.~(\ref{eq:integralbw}) in the following integral equation for the smooth
function $s({\bf k,r}_0)$ over the surface of the orifice
\begin{equation}
1=\frac{s({\bf k,r}_0)}{2|\ell_{\mathrm kz}|}+\int d\Omega_{k'}\, d{\bf
r'}_0\,G({\bf k,r}_0;{\bf k',r'}_0)s({\bf k',r'}_0), \label{eq:int}
\end{equation}
where $G({\bf k,r}_0;{\bf k',r'}_0)$ is non-singular part of the Green's
function~(\ref{eq:gb})
\begin{equation}
\label{eq:green}
G({\bf k,r}_0;{\bf k',r'}_0)=\frac{1}{32 \pi^{4}} \int d{\bf q}\,\frac{q\ell\,
e^{i{\bf q \cdot (r}_0-{\bf r'}_0)}}{(1+i{\bf q \cdot}
\bbox{\ell}_{\bf k})(q\ell-\arctan q\ell)(1+i{\bf q \cdot} 
\bbox{\ell}_{\bf k'})}.
\end{equation}
The distribution function $s({\bf k,r}_0)$ has two ${\bf k}$-space variables,
the polar and azimuthal angles $(\theta_k,\phi_k)$ of the vector ${\bf k}$ on
the Fermi surface, and the radius $r_0$ and azimuthal angle $\phi_0$ of the
point ${\bf r}_0$ on the orifice. Because of the cylindrical symmetry, $s({\bf
k,r}_0)$ does not depend separately on $\phi_k$, $\phi_0$, but only on their
difference $\phi_k-\phi_0$. This allows the expansion
\begin{equation}
\label{eq:expand}
s({\bf k,r}_0)=\sum_{LM}{s_{\mathrm LM}(r_0)Y_{\mathrm
LM}(\theta_{k},\phi_k)e^{-iM\phi_0}},
\end{equation}
and Eq.~(\ref{eq:int}) can now be rewritten as
\begin{eqnarray}
2\ell \cos \theta_{k} & = & \sum_{L'M'}s_{\mathrm L'M'}(r_0)Y_{\mathrm
 L'M'}(\theta_{k},\phi_{k})e^{-iM'\phi_0} \mathrm{sgn}\,(\cos \theta_k)
 \nonumber \\ \mbox{} & & + 2\ell\int d{\Omega_{k'}}\,d{\bf r'}_0\, G({\bf
 k,r}_0;{\bf k',r'}_0)\cos \theta_{k}\sum_{L'M'}s_{\mathrm L'M'}(r'_0)Y_{\mathrm
 L'M'}(\theta_{k'},\phi_{k'})e^{-iM'\phi'_0} \label{eq:int1}.
\end{eqnarray}
This four dimensional integral equation can be reduced to a system of coupled
one dimensional Fredholm integral equations of the second kind after it is
multiplied by $Y_{\mathrm LM}^{\ast}(\theta_k,\phi_k)e^{iM \phi_0}$ and
integrated over $\theta_k$, $\phi_k$ and $\phi_0$. We also use the following
identities
\begin{mathletters}
\begin{eqnarray}
\label{eq:sph}
Y_{\mathrm LM}(\theta,\phi)\, \cos \theta & = & g_1 Y_{\mathrm
L+1,M}(\theta,\phi)+ g_2 Y_{\mathrm L-1,M}(\theta,\phi),\\ g_1 & = &
\sqrt{\frac{(L-M+1)(L+M+1)}{(2L+1)(2L+3)}},\\ g_2 & = &
\sqrt{\frac{(L-M)(L+M)}{(2L-1)(2L+1)}},
\end{eqnarray}
\end{mathletters}
\begin{equation}
\label{eq:fl}
\frac{1}{4\pi} \int \frac{Y_{\mathrm{LM}}(\theta_k,\phi_k)}{1+i{\bf q \cdot}
\bbox{\ell}_{\bf k}}\, d\Omega_{k} =
i^{L}f_{L}(q\ell)Y_{\mathrm{LM}}(\theta_q,\phi_q),
\end{equation}
and
\begin{equation}
\int_{0}^{2\pi}e^{i{\bf qr}_0}e^{-iM\phi_0}\, d\phi_0=
\int_{0}^{2\pi}e^{iq_{\perp}r_0\cos (\phi_0-\phi_{q})}e^{-iM\phi_0}\, d\phi_0 =
2\pi i^{M}J_{M}(q_{\perp}r_0)e^{-iM\phi_q}, \label{eq:bes}
\end{equation}
where ${\bf q_{\perp}}$ is projection of ${\bf q=q_{z}+q_{\perp}}$ in the plane
of orifice and $J_{M}(z)$ is the Bessel function of the first kind.  For the
function $f_L(q\ell)$ in~(\ref{eq:fl}) we get the following expression
\begin{equation}
f_L(q \ell)=(-1)^L \int_0^{\infty} e^{-x} j_L(q \ell x)\, dx =\frac{(-i)^{-L}}{i
q\ell} Q_L(\frac{1}{i q \ell}),
\end{equation}
where $j_L(x)$ is spherical Bessel function and $Q_L(x)$ is Legendre function of
the second kind.  Explicit formulae for $f_L(x)$ are
\begin{mathletters}
\begin{eqnarray}
f_0(x) & = & \frac{\arctan x}{x}, \\ f_1(x) & = & \frac{-x+\arctan x}{x^2}, \\
f_2(x) & = & \frac{-3x+ (x^2+3) \arctan x}{2x^3}, \\ f_3(x) & = &
\frac{-\frac{4}{3}x^3-5x+ (5+3x^2) \arctan x}{2x^4}, \\ f_4(x) & = &
\frac{-\frac{55}{3}x^3-35x+(35+30x^2+3x^4) \arctan x}{8x^5}.
\end{eqnarray}
\end{mathletters}
The final form of the integral equation for $s_{\mathrm{LM}}(r_0)$ in the
expansion of $s({\bf k,r}_0)$ is
\begin{equation}
\label{eq:end}
4\ell \sqrt{\frac{\pi}{3}}\delta_{L1}\delta_{M0} =
\sum_{\mathrm{L'M'}}c_{\mathrm{LM,L'M'}}\delta_{\mathrm
MM'}s_{\mathrm{L'M'}}(r_0)+4\sum_{L'M'}\int_0^a r'_0\,dr'_0 \,
K_{\mathrm{LM,L'M'}}(r_0,r'_0) s_{\mathrm{L'M'}}(r'_0),
\end{equation}
where the kernel of equation is given by
\begin{eqnarray}
\label{eq:ker}
K_{\mathrm{LM,L'M'}}(r_0,r'_0) & = & i^{M'-M}(-1)^{M+M'} \int_{0}^{\infty} q^2
\, dq \int_0^{\pi} \sin \theta_q \, d\theta_q \, \frac{q\ell^2
f_{L'}(q\ell)Y_{\mathrm L'M'}(\theta_q)}{q\ell - \arctan q\ell}
 J_{M}(q r_0 \,\sin \theta_q)J_{M'}(qr'_0 \, \sin \theta_q) \nonumber \\ & &
\times (i^{L'+L+1}(-1)^{L+1}g_1f_{L+1}(q\ell)Y_{\mathrm L+1,M}(\theta_q)+
i^{L'+L-1}(-1)^{L-1}g_2f_{L-1}(q\ell)Y_{\mathrm L-1,M}(\theta_q)).
\end{eqnarray}
Kernel~(\ref{eq:ker}) does not depend on $\phi_q$ so that only $\theta$-part of
spherical harmonic $Y_{\mathrm LM}(\theta_q)$ (i.e. associated Legendre
polynomial) is integrated.  The kernel differs from zero only if $L+M$ has
parity different from $L'+M'$. This follows from the fact that the kernel is the
expectation value
\begin{eqnarray}
K_{\mathrm LM,L'M'}(r_0,r'_0) & = & \langle LMM|2\ell \, \cos \theta \, G({\bf
k,r}_0;{\bf k',r'}_0)|L'M'M' \rangle, \\ | LMM\rangle & = &
Y_{\mathrm{LM}}(\theta_k,\phi_k)e^{-iM\phi_0}
\end{eqnarray}
of an odd operator under inversion in the basis of functions $|LMM
\rangle$. Their parity is given by
\begin{equation}
P|LMM\rangle=(-1)^{L+M}|LMM\rangle.
\end{equation}
Exactly under this condition the kernel becomes a real quantity. This means that
nonzero $s_{\mathrm LM}(r_0)$ are real with the property
\begin{equation}
s_{\mathrm LM}(r_0)=(-1)^Ms_{\mathrm L,-M}(r_0),
\end{equation}
ensuring that $s({\bf k,r}_0)$ is real. The conductance is determined by the
$(L,M)=(0,0)$ function $s_{00}(r_0)$. The non-zero $s_{\mathrm{LM}}(r_0)$
coupled to it are selected by the condition that $L+M$ is even. This follows
from $s({\bf k,r}_0)$ being even under reflection in the plane of orifice. Under
this operation, $\cos \theta_k \rightarrow - \cos \theta_k$, but $\phi_k$,
$\phi_0$ are unchanged; this means that the expansion~(\ref{eq:expand}) contains
only terms with $L+M$ even.
 
The first term on the right hand side in~(\ref{eq:int1}) is determined by the
matrix element
\begin{equation}
\label{eq:clm}
c_{\mathrm{LM,L'M'}}=\int d\theta_k \, d\phi_k \,\sin \theta_k
Y_{\mathrm{LM}}^{\ast}(\theta_k,\phi_k)Y_{\mathrm{L'M'}}(\theta_k,\phi_k)\,
\mathrm{sgn}\,(\cos \theta_k),
\end{equation}
which is expectation value of $\mathrm{sgn}\,(\cos \theta_k)$ in the basis of
spherical harmonics. It is different from zero if $M=M'$ and $L-L'$ is odd. The
states must be of of different parity, as determined by $L$, because
$\mathrm{sgn}\,(\cos \theta_k)$ is odd under inversion.

\begin{figure}
\centerline{
\psfig{file=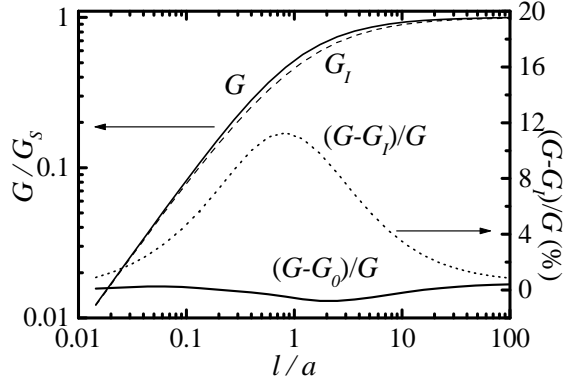,height=4.0in,angle=0}}
\vspace{-1.5in}
\caption{The conductance $G$ ($L=2$, $n=2$), normalized by the Sharvin
conductance $G_S$~(\ref{eq:shar}), plotted against the ratio $\ell/a$.  
It is compared to the naive interpolation formula $G_I$~(\ref{eq:intp}), 
and the plausible interpolation formula $G_0$~(\ref{eq:gamma}).}
\label{fig:fig3}
\end{figure}

The system of equations~(\ref{eq:end}) can be solved for all possible ratios of
$\ell/a$ by either discretizing variable $r_0$ or by expanding $s_{\mathrm
L'M'}(r_0)$ in terms of the polynomials in $r_0$
\begin{equation}
s_{\mathrm{LM}}(r_0)=\sum_n a_{\mathrm{nLM}} p_n(r_0),
\end{equation}
and performing integrations numerically. The polynomials $p_n(r_0)=\sum_{i=0}^n
c_i r^i$ are orthogonal with respect to the scalar product

\begin{equation}
\int_0^a r_0\,dr_0\,p_n(r_0)p_m(r_0)=\delta_{nm}.
\end{equation}
The first three polynomials are
\begin{mathletters}
\begin{eqnarray}
p_0(r_0) & = & \frac{\sqrt{2}}{a}, \\ p_1(r_0) & = & \frac{6r_0-4}{a
\sqrt{9a^2-16a+9}}, \\ p_2(r_0) & = &
\frac{10\sqrt{6}\left(r_0^2-\frac{6}{5}r_0+\frac{3}{10}\right)}
{a\sqrt{100a^4-288a^3+306a^2-144a+27}}.
\end{eqnarray}
\end{mathletters}
Each integral equation in the system~(\ref{eq:end}) then becomes the matrix
equation. Their introduction into inherent matrix structure of~(\ref{eq:end})
gives the matrix equation for either $s_{\mathrm LM}(r_0)$ at discretized $r_0$
or constants $a_{\mathrm{nLM}}$. Therefore, the constants $a_{\mathrm{nLM}}$
satisfy the following equation
    
\begin{mathletters}
\label{eq:varapp}
\begin{equation}
4\ell a\sqrt{\frac{\pi}{6}} \delta_{L1} \delta_{M0} \delta_{n0} =
\sum_{L'}c_{\mathrm{LM,L'M}}\, a_{\mathrm{nL'M}}+4\sum_{n'L'M'}
K_{\mathrm{nLM}}^{\mathrm{n'L'M'}} a_{\mathrm{n'L'M'}},
\end{equation}
\begin{eqnarray}
\label{eq:KnLM}
K_{\mathrm{nLM}}^{\mathrm{n'L'M'}} & = &i^{M'-M}(-1)^{M+M'} \int_{0}^{\infty}
q^2 \,dq \int_0^{\pi} \sin \theta_q \, d\theta_q \, \frac{q \ell^2
f_{L'}(q\ell)Y_{\mathrm L'M'}(\theta_q)}{q\ell-\arctan q\ell} j^n_M(qa \, \sin
\theta_q) j^{n'}_{M'}(qa \, \sin \theta_q) \nonumber \\ & & \times
(i^{L'+L+1}(-1)^{L+1}g_1f_{L+1}(q\ell)Y_{\mathrm L+1,M}(\theta_q)+
i^{L'+L-1}(-1)^{L-1}g_2f_{L-1}(q\ell)Y_{\mathrm L-1,M}(\theta_q)), \\
j^{n}_{M}(qa\, \sin \theta_q) & = &\int_0^a r_0\,dr_0\, p_n(r_0)J_M(q r_0 \,
\sin \theta_q),
\end{eqnarray}
\end{mathletters}
which simplifies using the following result
\begin{equation}
\label{eq:var2}
j^{n}_{M}(qa \, \sin \theta_q)= \sum_{i=0}^n c_i \frac {{a}^{2+M+i}(q \,\sin
\theta_q)^M {_1F_2}
(1+\frac{M}{2}+\frac{i}{2};2+\frac{M}{2}+\frac{i}{2},1+M;-\frac{1}{4}(qa \, \sin
\theta_q)^{2})}{2^{1+M}\left (1+\frac{M}{2}+\frac{i}{2}\right )\Gamma (1+M)},
\nonumber
\end{equation}
where $_1F_2(\alpha;\beta_1,\beta_2;z)$ is a hypergeometric function.  The
lowest order approximation for $s({\bf k,r}_0)$ is obtained by truncating the
polynomial expansion to zeroth order (i.e. constant---which is the space
dependence of Sharvin limit) and expansion in spherical harmonics to $L=0$. Then
the conductance is determined only by the constant $a_{000}$ following trivially
from~(\ref{eq:varapp})

\begin{equation}
\label{eq:exact}
G_{\mathrm lo}=\frac{N(0) \ell e^2 a^2 \pi}{\tau(3+K_{010}^{000})}, \nonumber
\end{equation}
where the lowest order part of the kernel $K_{010}^{000}$ depends on $\ell/a$,
\begin{equation}
K_{010}^{000} =  \frac{4\ell}{\pi} \int_0^{\infty} dq\, \int_0^{\pi}
d\theta_q\, \frac{\arctan q\ell}{q\ell-\arctan q\ell} \left( \frac{-3q\ell+ 
(q^2 \ell^2+3) \arctan q\ell}{2q^3 \ell^3}(1-3\cos^2 \theta_q)+\frac{\arctan
q\ell}{q\ell} \right ) \frac{(J_1(qa\, \sin
\theta_q))^2}{\sin \theta_q}.
\end{equation}
Further corrections are obtained by solving the matrix 
equation~(\ref{eq:varapp})
where the infinite matrix is approximated by its finite block. The matrix
elements $K_{\mathrm{nLM}}^{\mathrm{n'L'M'}}$~(\ref{eq:KnLM}) are tedious
to compute, but the conductance converges rapidly for large $n$ and $L$. 
On the other hand, the matrix elements $c_{\mathrm{LM,L'M'}}$~(\ref{eq:clm}) 
are easy to compute but the conductance converges slowly in the ballistic 
limit which is determined by this matrix elements. We keep the low order 
matrix elements 
$K_{\mathrm{nLM}}^{\mathrm{n'L'M'}}$ but go to high order in
$c_{\mathrm{LM,L'M'}}$. In practice we find that for \mbox{$c$-matrix} 
$L_{\mathrm max}=12$ is sufficient, whereas for \mbox{$K$-matrix} 
the approximation $L_{\mathrm max}=2$, $n_{\mathrm max}=2$ 
gives convergence to $1\%$.  
The conductance as a function of $\ell/a$ is shown
on Fig.~\ref{fig:fig3}. It is normalized to the Sharvin conductance,
i.e. conductance in the limit $\ell \gg a$, for which
\begin{eqnarray}
G({\bf k,r;k',r'}) & \rightarrow & 0, \nonumber \\ s({\bf k,r}) & = &
2|\ell_{\mathrm kz}|.
\end{eqnarray}
In the opposite (Maxwell) limit, when $\ell \ll a$, we have
\begin{mathletters}
\begin{eqnarray}
\frac{q\ell}{q\ell-\arctan q\ell} & = & \frac{3}{(q\ell)^2}+9/5+o((q\ell)^2), \\
G({\bf k,r ; k' , r'}) & \rightarrow & \frac{3}{32 \pi^4}\int d{\bf q}\,
\frac{e^{i{\bf q \cdot (r-r')}}}{(q\ell)^2} = \frac{3}{16\pi^2 \ell^2|{\bf
r-r'}|},
\end{eqnarray}
\end{mathletters}
which is the standard Green's function for the Poisson equation. The dependence
of the full Green's function~(\ref{eq:gb}) on ${\bf k}$ vector is reflection of
non-locality.  The conductance in the transition region from Maxwell to Sharvin
limit can be compared with the naive interpolation formula which approximates
resistance of orifice by the sum of Sharvin and Maxwell resistances
\begin{equation}
\label{eq:intp}
\frac{1}{G_{I}}=R_{I}=R_S \left (1+\frac{3\pi}{8} \frac{a}{\ell} \right ).
\end{equation}
Somewhat unexpectedly, naive interpolation formula $G_{I}$ deviates from our 
result for $G$ at most $11\%$ when $\ell/a \rightarrow 1$ as
shown on Fig.~\ref{fig:fig3}. We can also cast our lowest order approximation 
for the conductance~(\ref{eq:exact}) in an analogous form
as~(\ref{eq:intp})
\begin{equation}
\label{eq:num}	
\frac{1}{G_{\mathrm lo}}=R_S \left (\frac{3}{4}+ \frac{32}{3\pi^2} \gamma 
\frac{3\pi}{8} \frac{a}{\ell} \right ).
\end{equation}
The numerical coefficients in Eq.~(\ref{eq:num}) are not accurate in this
simplest approximation. Replacement of $3/4$ by $1$ and $32/(3\pi^2)$ by $1$
yields correct limiting values of $G$ and leads to a plausible interpolation
formula. It differs from Eq.~(\ref{eq:intp}) by the introduction of a factor
$\gamma$ which multiplies the Maxwell resistance

\begin{equation}
\label{eq:gamma}
\frac{1}{G_0}=R_S \left(1+ \gamma \frac{3\pi}{8} \frac{a}{\ell} \right ),
\end{equation}

\begin{equation}
\label{eq:gama}
\gamma = \frac{\pi \ell}{16a} K_{010}^{000}.
\end{equation}
This formula is compared to $G$ and $G_I$ on Fig.~\ref{fig:fig3}. It differs
from our most accurate calculation of $G$ by less then $1\%$. Therefore, 
for all practical purposes it can be used as an exact expression for the 
conductance in this geometry, and it is the main outcome of our work. 
The factor $\gamma$ is of order one and depends on the ratio $\ell/a$ as 
shown on Fig.~\ref{fig:fig2}. We also plot on Fig.~\ref{fig:fig2} 
Wexler's~\cite{wex} previous variational calculation, $\gamma_{\mathrm{Wex}}$.

In conclusion, we have calculated the conductance of the orifice in all
transport regimes, from the diffusive to the ballistic. The altered
version~(\ref{eq:gamma}) of the simplest approximate solution of our
theory~(\ref{eq:exact}) is already reasonably accurate. It shows  the 
microscopic theory correction to the naive interpolation formula 
(sum of Maxwell and Sharvin resistances) which is within 
$11\%$ different from the exact answer. Further corrections converge 
rapidly to an exact result. This analysis is of interest in any situation 
where the geometry of the sample can introduce additional resistances, 
like for example in transport in granular materials~\cite{hebard}.

This work was supported in part by the NSF grant no. DMR 9725037.
%


\end{document}